\input{aipcheck}

\documentclass[
    ,final            
  ]
  {aipproc}

\layoutstyle{6x9}

\begin{document}

	\title{Nuclear correction factors from neutrino DIS}


	\author{K.~Kova\v{r}\'{\i}k}{
	  address={Institute for Theoretical Physics, Karlsruhe Institute of Technology, Karlsruhe, D-76128,Germany}
	}

\begin{abstract}
	Neutrino Deep Inelastic Scattering (DIS) on nuclei is an essential process to constrain the strange quark parton 
	distribution functions (PDF) in the proton. The critical component on the way to using the neutrino DIS data in a proton PDF 
	analysis is understanding the nuclear effects in parton distribution functions. We parametrize these effects by nuclear 
	parton distribution functions (NPDF) and we use this framework to analyze the consistency of neutrino DIS data with 
	other nuclear data. 
\end{abstract}

\maketitle


\section{Introduction}
Any prediction for a process measured at a hadron collider such as the LHC, involves necessarily 
parton distribution functions (PDFs). Because of how crucial the knowledge of PDFs is, there are many groups 
that perform and update global analyses of PDFs protons \cite{Ball:2009mk, 
Martin:2009iq, Nadolsky:2008zw} and nuclei \cite{Hirai:2007sx, 
Eskola:2009uj}. Although not often emphasized, nuclear effects are present also in the 
proton PDFs analysis as a number of experimental data is 
taken on nuclear targets. Mostly though, the nuclear targets used in the proton analysis, are made of 
light nuclei where nuclear effects are generally 
small. An important exception is the neutrino DIS data which is taken on heavy nuclei such as iron or 
lead and is sensitive to the strange quark content of the proton. A knowledge of the strange quark PDF 
has an influence on precise measurements at the LHC such as $W$- or $Z$-boson production.

In order to include the neutrino DIS data in a global fit to determine proton PDF, we 
have to apply a nuclear correction factor. The nuclear correction factor can be obtained either from a 
specific model of nuclear interactions or from an analysis of nuclear parton 
distribution functions (NPDF) based on experimental data.

Here, we present a framework for a global analysis of nuclear PDFs at next-to-leading order in QCD 
closely related to the CTEQ framework for proton PDFs. We analyze and compare the nuclear correction 
factor obtained from the usual charged lepton DIS and 
Drell-Yan (DY) data to the one from the neutrino DIS data mainly from the NuTeV experiment.   
\section{Nuclear PDF}
To determine parton distribution functions from experimental data, we parametrize the $x$-dependence of PDFs 
at an input scale $Q_0$ and use the DGLAP equations to change the scale of experimental data to $Q_0$ in order 
to perform a fit to the data. The global NPDF framework, we use to analyze charged lepton DIS and DY data and 
neutrino DIS data, was introduced in \cite{Schienbein:2009kk}. The parameterizations of the nuclear parton distributions 
of partons in bound protons at the input scale of $Q_0=1.3 {\rm GeV}$	
\begin{equation}
x\, f_{k}(x,Q_{0}) = c_{0}x^{c_{1}}(1-x)^{c_{2}}e^{c_{3}x}(1+e^{c_{4}}x)^{c_{5}}\,,\label{eq:input1}
\end{equation}
where $k=u_{v},d_{v},g,\bar{u}+\bar{d},s,\bar{s}$ and
\begin{equation}	
\bar{d}(x,Q_{0})/\bar{u}(x,Q_{0}) = c_{0}x^{c_{1}}(1-x)^{c_{2}}+(1+c_{3}x)(1-x)^{c_{4}}\,,\label{eq:input2} 
\end{equation}
are a generalization of the parton parameterizations in free protons used in the CTEQ proton analysis \cite{Pumplin:2002vw}. 
To account for different nuclear targets, the coefficients $c_k$ are made to be functions of the nucleon number $A$
\begin{equation}
c_{k}\to c_{k}(A)\equiv c_{k,0}+c_{k,1}\left(1-A^{-c_{k,2}}\right),\ k=\{1,\ldots,5\}\,.\label{eq:Adep}
\end{equation}
The proton PDF in this framework are obtained as a limit $A\rightarrow 1$ and are held fixed at values obtained in the analysis 
\cite{Pumplin:2002vw}. From the input distributions, we can construct the PDFs for a general $(A,Z)$-nucleus 
\begin{equation}
f_{i}^{(A,Z)}(x,Q)=\frac{Z}{A}\ f_{i}^{p/A}(x,Q)+\frac{(A-Z)}{A}\ f_{i}^{n/A}(x,Q),\label{eq:pdf}
\end{equation}
where we relate the distributions of a bound neutron, $f_{i}^{n/A}(x,Q)$, to those of a proton by isospin symmetry. 

In the analysis, the same standard kinematic cuts $Q>2 {\rm GeV}$ and $W>3.5 {\rm GeV}$ were applied as in 
\cite{Pumplin:2002vw} and we obtain a fit with $\chi^{2}/{\rm dof}$ of 0.946 to 708 data points with 32 free parameters (for 
further details see \cite{Schienbein:2009kk}). 
%

The nuclear effects extracted in the form of NPDF are usually presented in the form of nuclear correction 
factors. We discuss two nuclear correction factors in the following where both are related either to the DIS structure 
function $F_2$ in the charged-current (CC) $\nu A$ process
\begin{equation}
R_{CC}^\nu(F_2;x,Q^2)\simeq \frac{d^A+\bar{u}^A+\ldots}{d^{A,0}+\bar{u}^{A,0}+\ldots}\,,
\label{eq:rcc}
\end{equation}
or to the DIS structure function $F_2$ in the neutral-current (NC) $l^\pm A$ process
\begin{eqnarray}
&& R_{NC}^{e,\mu}(F_2;x,Q^2)\simeq
\frac{[d^A + \bar{d}^A + \ldots]+ 4 [u^A + \bar{u}^A+\ldots]}{[d^{A,0} + \bar{d}^{A,0} + \ldots]
+4 [u^{A,0} + \bar{u}^{A,0}+\ldots]}\,.
\label{eq:rnc}
\end{eqnarray}
The superscript $`0'$ stands for using the free nucleon PDFs $f_i^{p,n}(x,Q)$ in Eq.\ (\ref{eq:pdf}). 

In Fig.~\ref{fig:f2-1} (solid line), we show how the result of our global analysis of charged lepton data translates into these 
nuclear correction factors and how it compares with experimental data. As first observed in \cite{Schienbein:2007fs}, the 
$R_{CC}^\nu(F_2;x,Q^2)$ correction factor calculated using Eq.~\ref{eq:rcc} with parton densities from the fit to the charged 
lepton nuclear data, does not describe the NuTeV data well which raises the question if including neutrino 
DIS data in the global analysis corrects this behavior without spoiling the $R_{NC}^{e,\mu}(F_2;x,Q^2)$ correction factor which 
fits the charged lepton DIS and DY data well.
\section{Neutrino DIS}
In order to analyze the apparent discrepancy between the nuclear correction factor $R_{CC}^\nu(F_2;x,Q^2)$ from the 
fit to charged lepton data and the neutrino charged current DIS data, we have 
set up a global analysis where we used exclusively the neutrino DIS cross-section data coming from NuTeV and Chorus experiments 
taken on iron and lead respectively. Here we applied the same kinematic cuts as in the first analysis of the charged lepton data 
and we obtain a fit to 3134 neutrino DIS cross-section data points with $\chi^{2}/{\rm dof}$ of 1.33 with 34 free parameters (for 
further details see \cite{Kovarik:2010uv}).
\begin{table}
\begin{tabular}{lccc}
\hline 
$w$  & $ \chi^{2}_{l^{\pm}A}$ (/pt)  & $\chi^{2}_{\nu A}$ (/pt)  & total $\chi^{2}$(/pt)\tabularnewline
\hline 
$0$    & 638 (0.90)   & -  & 638 (0.90) \tabularnewline
$1/7$   & 645 (0.91)    & 4710 (1.50)  & 5355 (1.39) \tabularnewline
$1/2$    & 680 (0.96)   & 4405 (1.40)  & 5085 (1.32) \tabularnewline
$1$   & 736 (1.04)      & 4277 (1.36)  & 5014 (1.30) \tabularnewline
$\infty$  & -   & 4192 (1.33)  & 4192 (1.33) \tabularnewline
\hline
\end{tabular}
\caption{Summary table of a family of compromise fits. \label{tab:compr} }
\end{table}
A global fit to neutrino DIS data describes the data for the charged current nuclear correction factor 
$R_{CC}^\nu(F_2;x,Q^2)$ well and does a poor job to describe the neutral current correction factor especially at low and 
intermediate Bjorken $x$. In order to find nuclear correction factors in agreement with both charged lepton and neutrino 
data, we perform a combined global analysis of both data sets. We introduce an additional parameter in our analysis, 
the weight of the neutrino data set $w$, which should prevent the neutrino DIS data set to dominate the global fit based
only on the number of data points taken. The weight $w$ enters the calculation of the combined $\chi^2$ as
\begin{equation}
\chi^{2}=\sum_{l^{\pm}A\ {\rm data}}\chi_{i}^{2}\ +\!\!\sum_{\nu A\ {\rm data}}w\,\chi_{i}^{2}\ ,\label{eq:chi2}
\end{equation}
\begin{figure}[t]
\includegraphics[width=.4\textwidth]{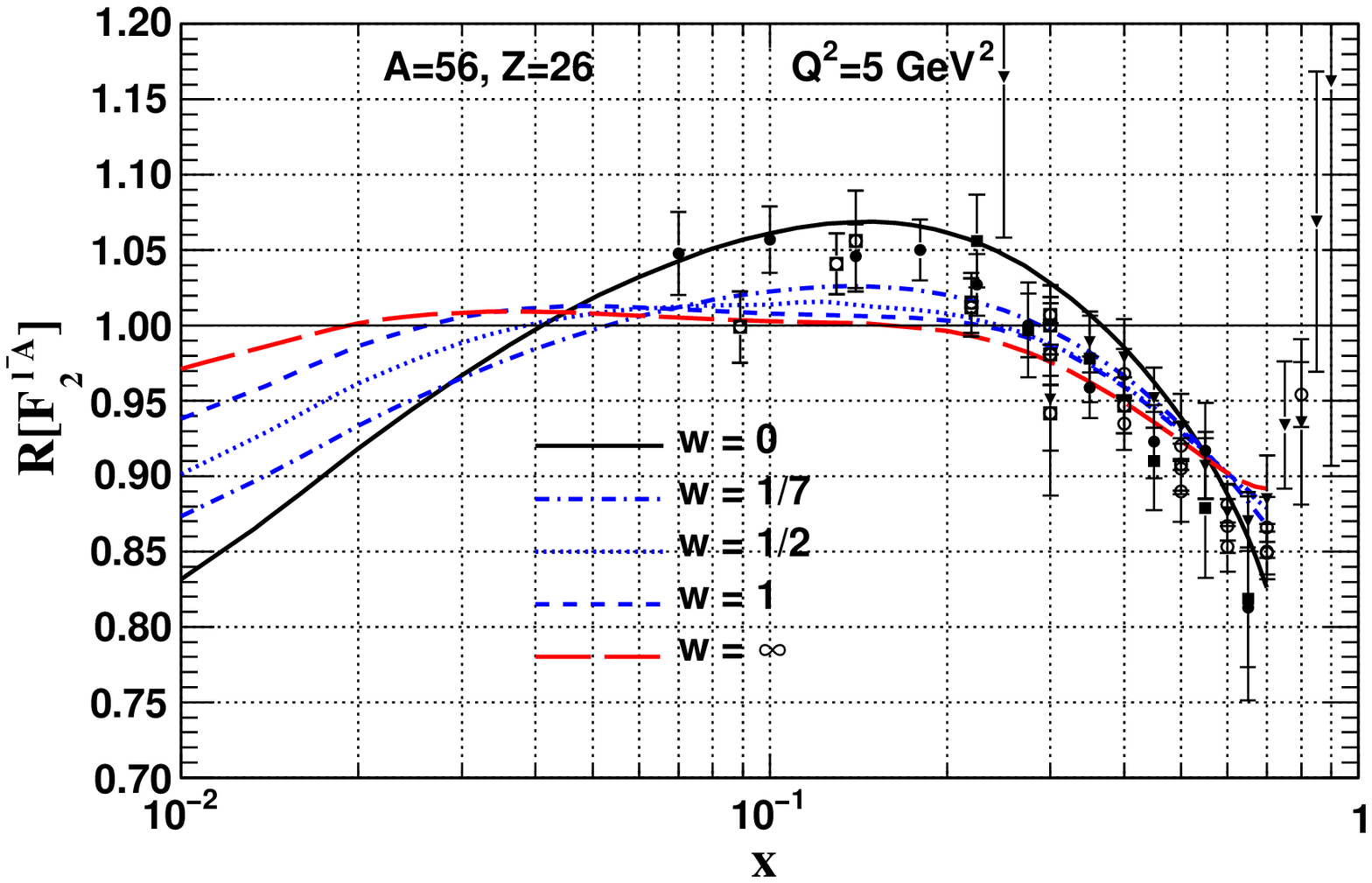}
\includegraphics[width=.4\textwidth]{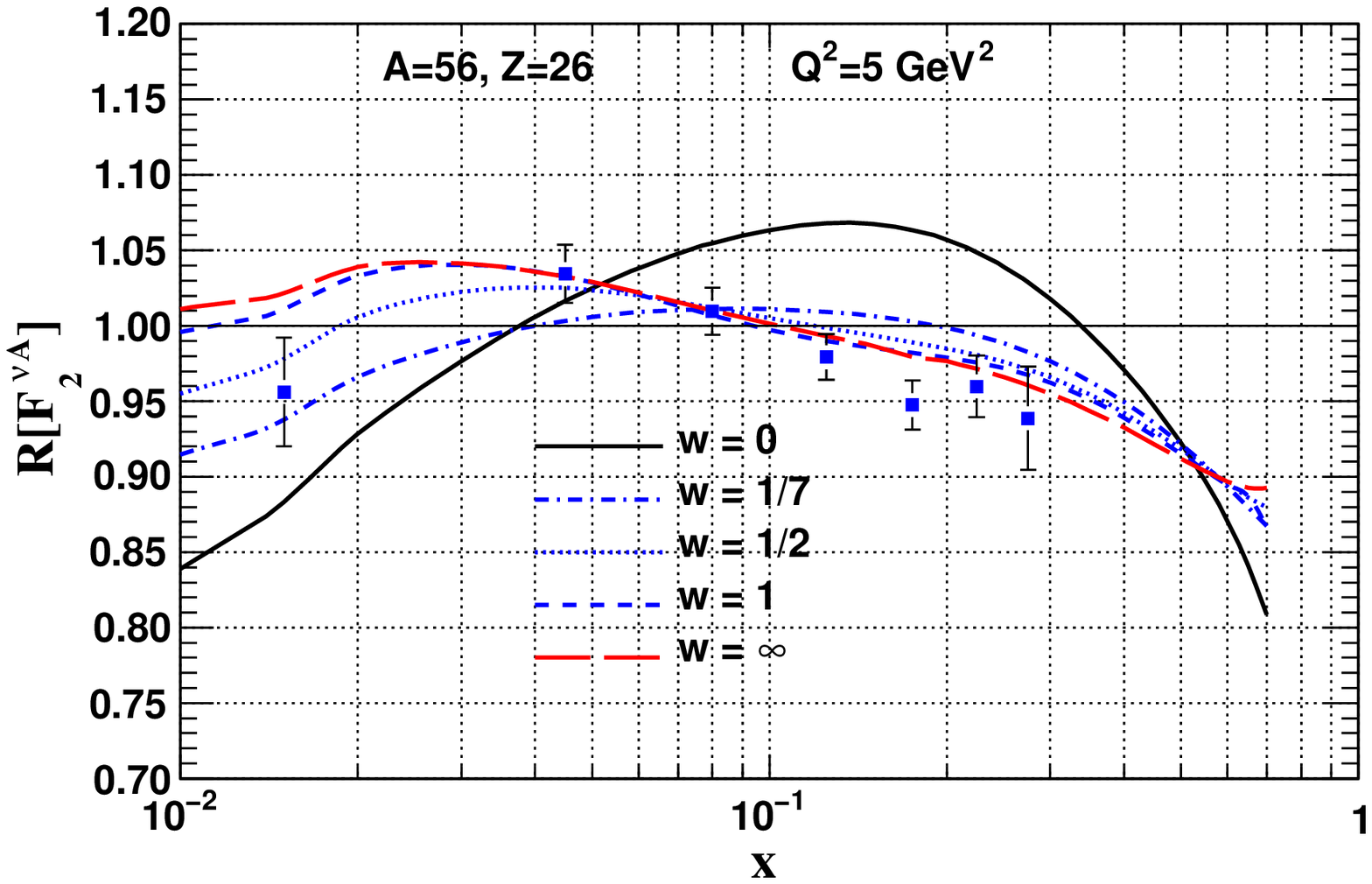}
  \caption{Nuclear correction factors $R_{NC}^{e,\mu}(F_2;x,Q^2)$ (left) and $R_{CC}^{e,\mu}(F_2;x,Q^2)$ (right) for compromise fits with different weights of the neutrino DIS data at the scale $Q^2=5 {\rm GeV}^2$.}\label{fig:f2-1}
\end{figure}
and it can be used to interpolate between the two different global fits ($w=0$ results in the fit to charged lepton data only 
and $w=\infty$ stands symbolically for the fit only to neutrino data). Varying the weight $w$, we try to find a compromise fit which 
would describe both charged lepton and neutrino data well. We list the resulting $\chi^2$ for the compromise fits with weights 
$w=0,1/7,1/2,1,\infty$ in Tab.~\ref{tab:compr} and we show the nuclear correction factors 
$R_{NC}^{e,\mu}(F_2;x,Q^2)$ and $R_{CC}^\nu(F_2;x,Q^2)$ for the same family of compromise fits in Fig.~\ref{fig:f2-1}.

In order to decide on how well the compromise fits describe the data we use the $\chi^2$ goodness-of-fit criterion used in 
\cite{Stump:2001gu,Martin:2009iq}. We consider a fit a good compromise if its $\chi^2$ for both data subsets, the charged lepton DIS and DY 
data and the neutrino DIS data, is within 90\% confidence level of the fits to only charged lepton or neutrino data. 

We define the 90\% percentile $\xi_{90}$ used to define the 90\% confidence level, by
\begin{equation}\label{xi90}
	\int_0^{\xi_{90}}P(\chi^2,N)d\chi^2 = 0.90\,,
\end{equation}
where $N$ is the number of degrees of freedom and $P(\chi^2, N)=\frac{(\chi^2)^{N/2-1}e^{-\chi^2/2}}{2^{N/2}\Gamma(N/2)}$ is the probability distribution. We can assign a 90\% confidence level error band to the $\chi^2$ of the fits to the charged lepton DIS and DY data 
and to the neutrino DIS data 
\begin{equation}\label{lnuA90}
	\chi^2_{l^\pm A} = 638+ 45.6,\qquad \chi^2_{\nu A} = 4192+ 138.
\end{equation}
Comparing the results of the compromise fits with different weights, listed in Tab.~\ref{tab:compr}, we see that none of the compromise fits 
are compatible with both 90\% confidence level limits given in Eq.\ref{lnuA90}. As detailed in \cite{Kovarik:2010uv}, not even relaxing 
the condition to compare against the 99\% confidence level limit helps to finding a suitable compromise fit. Moreover, we show in 
\cite{Kovarik:2010uv} that the effect is related to the precise neutrino DIS data from NuTeV.
\section{Conclusion}
After performing a thorough global NPDF analysis of the combined charged lepton and neutrino data, we find that there is no good compromise 
description of both the data sets simultaneously. The differences are most pronounced in the low and intermediate $x$ regions where the 
neutrino DIS (NuTeV) do not show a strong shadowing effect as the charged lepton data do. The inability to describe all data by one consistent 
framework indicates the existence of non-universal nuclear effects or unexpectedly large higher-twist effects.  



\bibliographystyle{aipproc}   

\bibliography{npdf}

%

\end{document}